\documentclass[10pt,twocolumn,letterpaper]{article}

\usepackage[pagenumbers]{cvpr} 

\usepackage[dvipsnames]{xcolor}
\usepackage{tabularx, booktabs}
\usepackage{tikz}

\usepackage{tabularx}
\usepackage{booktabs}
\usepackage{adjustbox}
\usepackage{tabularx}
\usepackage{booktabs}
\usepackage{graphicx}
\usepackage{multirow}
\usepackage{indentfirst}

\definecolor{cvprblue}{rgb}{0.21,0.49,0.74}

\usepackage[pagebackref,breaklinks,colorlinks,citecolor=cvprblue]{hyperref}

\def\paperID{43} 
\def\confName{CVPR}
\def\confYear{2024}

\title{Bracketing Image Restoration and Enhancement with High-Low Frequency Decomposition}

\author{Genggeng Chen$^{1\dag}$ \ \ \ \ Kexin Dai$^{2\dag}$ \ \ \ \ Kangzhen Yang$^{2}$ \ \ \ \ Tao Hu$^{1,2}$ \ \ \ \ Xiangyu Chen$^{3}$ \ \ \\
\ \ Yongqing Yang$^{4}$ \ \ \ \ Wei Dong$^{1}$  \ \ \ \ Peng Wu$^{2}$  \ \ \ \ Yanning Zhang$^{2}$\ \ \ \ Qingsen Yan$^{2}$\thanks{Corresponding author. $\dag$~The first two authors contributed equally to this work.
This work was partially supported by NSFC (62301432,62306240), NSBRPS (2023-JC-QN-0685, QCYRCXM-2023-057). 
}\\
$^{1}$Xi’an University of Architecture and Technology\ $^{2}$Northwestern Polytechnical University\\ $^{3}$University of Macau\ $^{4}$Xi'an Institute of Optics and Precision Mechanics of CAS \\
\url{https://github.com/chengeng0613/HLNet}}

\begin{document}
\maketitle
\begin{abstract}
\label{sec:Abstract}
In real-world scenarios, due to a series of image degradations, obtaining high-quality, clear content photos is challenging. While significant progress has been made in synthesizing high-quality images, previous methods for image restoration and enhancement often overlooked the characteristics of different degradations. They applied the same structure to address various types of degradation, resulting in less-than-ideal restoration outcomes. Inspired by the notion that high/low frequency information is applicable to different degradations, we introduce HLNet, a Bracketing Image Restoration and Enhancement method based on high-low frequency decomposition. Specifically, we employ two modules for feature extraction: shared weight modules and non-shared weight modules. In the shared weight modules, we use SCConv to extract common features from different degradations. In the non-shared weight modules, we introduce the High-Low Frequency Decomposition Block (HLFDB), which employs different methods to handle high-low frequency information, enabling the model to address different degradations more effectively. Compared to other networks, our method takes into account the characteristics of different degradations, thus achieving higher-quality image restoration.
\end{abstract}
    
\section{Introduction}
\label{sec:Introduction}

In real-world scenarios, the presence of various image degradations makes it challenging to capture high-quality, clear content photos. Low exposure can lead to increased noise, especially in dark areas, potentially causing loss of detail. Similarly, bright areas in high-exposure images may lose detail due to overexposure. Despite numerous single-image restoration methods proposed, such as denoising\cite{abdelhamed2020ntire,brooks2019unprocessing,guo2019toward,li2023ntire,zamir2020cycleisp,zhang2017beyond}, deblurring\cite{cho2021rethinking,mao2023intriguing,nah2017deep,tao2018scale,zamir2020cycleisp}, super-resolution\cite{dong2015image,ledig2017photo,liang2021swinir} and high dynamic range image reconstruction\cite{eilertsen2017hdr,lecouat2022high}, their performance is constrained by the insufficient information present in single images. 

\begin{figure}
    \centering
    \includegraphics[width=1\linewidth]{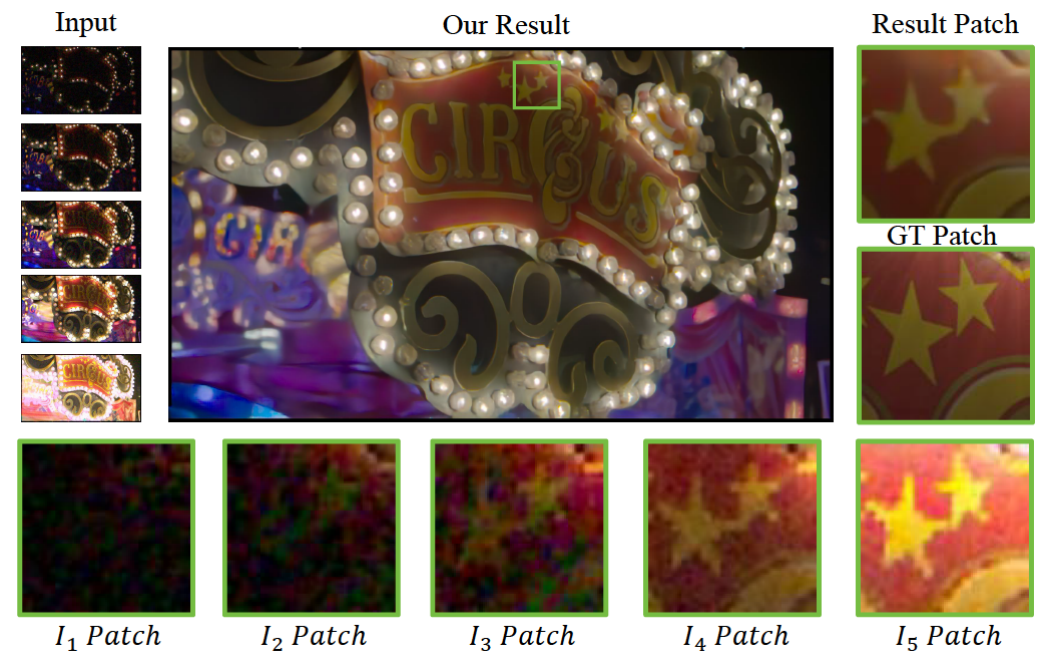}
    \caption{Our HLNet excels in restoring image details and texture, effectively enhancing the edge details of the stars in the figure.}
    \label{fig:Intro}
    \vspace{-1.0em}
\end{figure}
Due to the limitations of single image restoration and enhancement, such as insufficient information and exposure time impact, an increasing number of methods are utilizing multiple frames for restoration. Burst image restoration methods\cite{dudhane2022burst,dudhane2023burstormer} use multiple consecutive frames for super-resolution and denoising, while multi-exposure HDR imaging\cite{kalantari2017deep,liu2022ghost,niu2021hdr,prabhakar2020towards,yan2019attention,yan2020deep,yan2023towards,yan2022dual,yan2023smae,yan2023unified} reconstructs HDR images from LDR images with different exposures. However, these methods only consider a single degradation scenario, overlooking other degradation situations. In recent times, TMRNet\cite{zhang2024bracketing} has proposed a viable solution in the framework design for unified image restoration and enhancement tasks. It takes multi-exposure images as input and progressively blends non-reference frames with reference frames. 
Although they have considered the issues of commonality and specificity of different degradations, using shared weight modules and non-shared weight modules, they used the same structure when dealing with the characteristics of different degradations, ignoring some characteristics of different degradations.

To address the aforementioned issues, we propose HLNet, which takes into account the characteristics of various degradations. Similar to TMRNet, our model utilizes both shared-weight modules and non-shared-weight modules to extract features. In the shared-weight module, we introduce the Spatial Channel Enhancement Block (SCEB), which utilizes SCConv to simultaneously consider spatial and channel information, effectively extracting common features of different degradations. In the non-shared weight module, different degradations are suited to be processed with different frequency information. For instance, denoising and deblurring usually require enhancing high-frequency information to restore image details and textures, super-resolution requires adding high-frequency details on top of restoring low-frequency information, while HDR reconstruction requires capturing low-frequency information from different frames. Hence, we propose the High-Low-Frequency Decomposition Block (HLFDB). 

Specifically, in the HLFDB, high-frequency features can capture detailed local information, making them more suitable for denoising and deblurring enhancement. Therefore, we extract local feature maps through multiple convolutional blocks and enhance high-frequency details among multiple frames via dense connection mechanisms. Low-frequency features can capture most structural information and global features in the image, making them more suitable for super-resolution and multi-exposure HDR reconstruction tasks. Hence, we employ multi-level channel self-attention to learn long-range dependencies and utilize a scale-wise feature fusion method based on wavelet transform to avoid the loss of structural information caused by downsampling. Therefore, our model fully considers the characteristics of different degradations when it comes to image restoration and enhancement, enabling better performance in unified image restoration and enhancement tasks.

The main contributions are summarized as follows:
\begin{itemize}
    \item We propose the SCEB, which utilizes SCConv to simultaneously consider spatial and channel information, effectively extracting common features of different degradations.
    \item We have proposed HLFDB, which fully considers that different degradations require different high and low-frequency information, effectively addressing the challenge of simultaneously restoring these degradations.
    \item HLNet outperformed previous state-of-the-art (SOTA) models in both metrics and visual quality, achieved fourth place in track 2 of the Bracketing Image Restoration and Enhancement Challenge.
\end{itemize}
\section{Related Work}
\label{sec:Related work}
\paragraph{Burst Image Restoration}
Burst image refers to a series of images captured in rapid succession over a short period of time. In this sequence of images, there may be slight variations, such as camera movement, object motion, or changes in lighting conditions. Burst image restoration typically involves several main categories: denoising, deblurring, super-resolution. Many methods focused on denosing have been widely studied in literature \cite{godard2017deep,guo2022differentiable,Marinc_2019,mildenhall2018burst,xia2020basis,xu2019learning,Guo_2021}. In certain methodologies, recurrent fully convolutional deep neural network \cite{godard2017deep} are employed, while some others opt for spatially varying kernel estimation \cite{mildenhall2018burst,Marinc_2019,xia2020basis,xu2019learning}. Notably, in \cite{Guo_2021},  offset estimation is additionally leveraged to address challenges arising from substantial object motion. In \cite{guo2022differentiable} proposed a two-stage training scheme, sequentially aligning at the patch level and pixel level, to achieve robust alignment between image frames.

Some methods explore the potential of burst image super-resolution \cite{wei2023towards,Dudhane_2022_CVPR,Mehta_2022_CVPR,Bhat_2021_CVPR}, Both \cite{wei2023towards} and \cite{Mehta_2022_CVPR} employ a transformer architecture; however, \cite{wei2023towards} diverges by omitting pixel-wise alignment, opting instead for a straightforward homography alignment based on structural geometry. \cite{Dudhane_2022_CVPR}'s central concept revolves around generating a collection of pseudo-burst features, seamlessly amalgamating complementary information from all input burst frames for effective information exchange.

\paragraph{Multi-frame HDR Restoration}
Multi-frame HDR restoration involves creating High Dynamic Range (HDR) images from multiple low dynamic range (LDR) frames.
In classic HDR restoration methods, Debevec \etal \cite{debevec2023recovering} first proposed the idea of merging multiple LDR images into a single HDR image. Subsequently, many methods have employed ways to align other frames to the reference frame, including optical flow, energy optimization, rank minimization, and others. \citet{zhang2011gradient} recalibrated motion region weights using image gradients. \citet{bogoni2000extending} computed flow vectors for alignment purposes.
Sen \etal \cite{sen2012robust} employed a patch-based energy minimization method to optimize subsequent alignment and reconstruction.
However, these methods perform poorly when faced with significant movement of foreground objects or excessive pixel loss in over-exposed or under-exposed areas.

\begin{figure*}[t]
    \centering
    \includegraphics[width=1\linewidth]{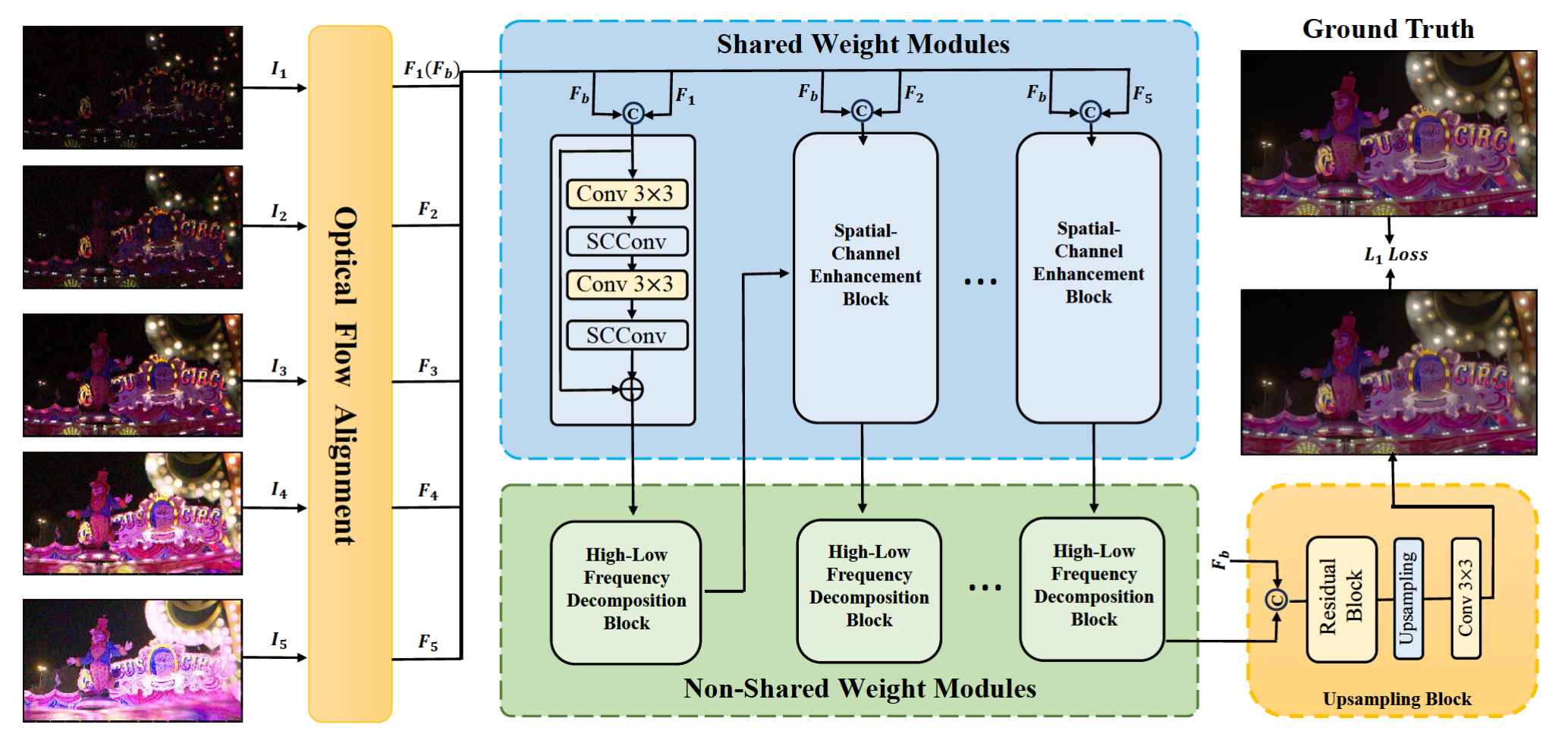}
    \caption{Overview of HLNet. In HLNet, feature alignment is performed first, followed by the gradual feeding of each frame into the network. The feature extraction stage consists of shared weight modules and non-shared weight modules. Each frame needs to pass through both modules, first through the shared weight module and then through the non-shared weight module. }
    \label{fig:Overview}
    \vspace{-1.0em}
\end{figure*}

With the development of deep learning, deep methods have also been applied to the field of multi-frame HDR fusion. an \etal \cite{yan2019attention}proposed a method of attention-guided image fusion, which reduces the presence of ghosting artifacts. SCTNet \cite{tel2023alignment} utilizes spatial attention and channel attention modules, aiming to simultaneously leverage dynamic and static contextual information for better image generation. However, these methods can only handle individual degraded images.

\section{Proposed Method}
\label{sec:Proposed method}

In this study, inspired by the collaborative potential of multi-exposure images and the applicability of high-low frequency information to different degradations, we propose a method for synthesizing and enhancing images through high-low frequency decomposition. The objective is to achieve clear, high dynamic range, and high-resolution images. Specifically, our input consists of five different-exposure raw images, denoted as $\{R_1, R_2, R_3,R_4, R_5\}$, which are then processed.

Initially, the number of multiple exposure images we input is 5, and we label the raw image captured with exposure time $\Delta t_i$ as $Y_i$, where $i \in \{1, 2, ..., 5\}$ and $\Delta t_i < \Delta t_{i+1}$. Subsequently, adhering to the guidelines established by multi-exposure HDR reconstruction methodologies\cite{yan2023towards,yan2019attention,yan2022dual,hu2024generating}, we merge the gamma-transformed image $Y_r$ with the original image $Y_i$ through concatenation, resulting in $F_i$, where $i \in \{1, 2, ..., 5\}$. This process can be formulated as follows: 
\begin{equation}
Y_{i} = \left( \frac{R_{i}}{\Delta t_{i} / \Delta t_{1}} \right),
Y_{r} = \left( \frac{R_{i}}{\Delta t_{i} / \Delta t_{1}} \right)^{\gamma},
\end{equation}
\begin{equation}
F_i=\text{Concat}(Y_{i},Y_{r}),
\end{equation}
where \(\gamma\) represents the gamma correction parameter and is generally set to $\frac{1}{2.2}$. Finally, we feed these concatenated images into the HLNet model. The resulting image is denoted as $\hat{H}$, and the process can be represented as:
\begin{equation}
\hat{H} = f(F_1, F_2, F_3, F_4, F_5; \theta),
\end{equation}
where $f (\cdot)$ represents the imaging function, and $\theta$ refers to the network's parameters.
\subsection{Overview of the HLNet}
The main framework of our model is illustrated in Fig.~\ref{fig:Overview}. Feature alignment is conducted first, inspired by TMRNet, where both shared-weight modules and non-shared-weight modules are utilized for feature extraction. Finally, upsampling is performed to obtain larger-sized images.

During the feature alignment stage, we select the first frame of the five input images as the reference frame and use optical flow\cite{bhat2021deep} to warp the features of other frames to align with the reference frame. Then, we further align the features using Deformable Convolutional Networks (DCN)\cite{dai2017deformable}.

In the feature extraction stage, we utilize both shared-weight modules and non-shared-weight modules. This is because in burst and video restoration tasks, some degradation types among multiple input frames are typically similar. Therefore, we employ shared-weight modules here. Shared weights not only enhance the model's generalization ability, enabling it to learn more universal feature representations, but also reduce the model's parameter count. By contrast, in TMRNet, only regular convolutions were used, without fully considering image details and contextual information. Hence, we introduce the Spatial-Channel Enhancement Block (SCEB), employing SCConv to simultaneously consider spatial and channel information, thus effectively capturing image details and contextual information. Additionally, relying solely on shared-weight modules is insufficient because other degradations are varying. For instance, differences in exposure time and image blur exist. Therefore, we introduce the High-Low Frequency Decomposition Block (HLFDB) to learn the specificity of different degradation types. This block explicitly separates the high and low-frequency information of the feature and processes them differently based on their characteristics.

At this stage, each aligned frame image $F_i$ , where $i \in \{1, 2, ..., 5\}$, is sequentially inputted into the network. Each frame image first passes through a shared-weight module, followed by a non-shared-weight module. Simultaneously, along with inputting each frame image, we also input the reference frame $F_b$($F_1$) and the output of the previous frame through the non-shared-weight module. This process can be expressed using the following formula:
\begin{equation}
F_{i+1_{out}}=\text{HLFDB}(\text{SCEB}(\text{Concat}(F_{i+1},F_{b},F_{i_{out}}))),
\end{equation}
$F_{i+1_{out}}$ denotes the output of the $(i+1)$-th non-shared-weight module. $\text{SCEB}(\cdot)$ represents the feature map after passing through the Spatial-Channel Enhancement Block, and $\text{HLFDB}(\cdot)$ represents the feature map after passing through the High-Low Frequency Decomposition Block.

In the final stage, the obtained feature maps are upsampled to map low-resolution images or feature maps to high resolution. This process is achieved through skip connections, feature fusion, and multiple upsampling operations.

\subsection{Spatial-Channel Enhancement Block}
In the Spatial-Channel Enhancement Block, we utilize SCConv\cite{li2023scconv}, a module that integrates spatial and channel dimensions in convolutional operations, enabling more effective capture of image details and contextual information, thus facilitating the extraction of common features across different degradations. In our module, we alternate between regular convolutions and SCConv, augmented with residual connections. This is because regular convolutions focus on capturing local spatial features, effectively extracting texture and shape information from the image. Meanwhile, SCConv considers the relationship between spatial and channel information, enabling a more comprehensive understanding of contextual information and inter-channel dependencies. By alternating between the two, we ensure that the model not only extracts powerful local features but also comprehends their importance in the global context, thus achieving a richer feature representation.The process can be expressed by the following formula:
\begin{equation}
F_{out}=F+\text{SCConv}(\text{Conv}(\text{SCConv}(\text{Conv}(F)))),
\end{equation}
where $F_{out}$ represents the output of the Spatial-Channel Enhancement Block, $\text{Conv}(\cdot)$ denotes 3 ${\times}$ 3 convolution, and $\text{SCConv}(\cdot)$ denotes Spatial-Channel Convolution.
\subsection{High-Low Frequency Decomposition Block}
\begin{figure}
    
    \centering
    \includegraphics[width=1\linewidth]{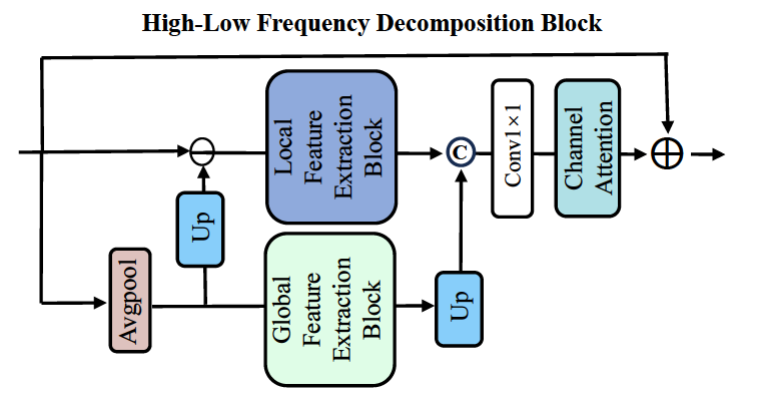}
    \caption{The architecture of the proposed High-Low Frequency Decomposition Block starts by using Average pooling to extract the low-frequency information from the features. Then, subtracting this low-frequency information from the overall features yields the high-frequency information.}
    \label{fig:HL}
    \vspace{-1.5em}
\end{figure}

To address the issue of the varying characteristics of different degradations during image restoration, we were inspired by ESRT\cite{lu2022transformer} and recognized the applicability of high-low frequency information to different degradation types. High-frequency information typically reflects local details in images, aiding in the restoration of fine image details, which is beneficial for tasks like deblurring and denoising. Meanwhile, low-frequency information generally represents the overall structure of the image, assisting in the restoration of image backgrounds and outlines, which is helpful for tasks like super-resolution reconstruction and multi-frame HDR reconstruction. Therefore, we separate the high and low-frequency information of the image and select different methods for feature extraction based on their characteristics.

As shown in Fig.~\ref{fig:HL}, we choose to use Avgpooling to obtain the low-frequency information of the features. Then, by subtracting the low-frequency information from the overall features, we obtain the high-frequency information of the features. Subsequently, we employ different methods to extract features from these two types of information. This process can be represented by the following formula:
\begin{equation}
F_{\text{low}}=\text{Avgpool}(F),
\end{equation}
\begin{equation}
F'=\text{Upsampling}(F_{\text{low}}),
F_{\text{high}}=F - F',
\end{equation}
where $\text{Avgpool}(\cdot)$ represents average pooling, and $\text{Upsampling}(\cdot)$ denotes bilinear interpolation upsampling. $F_{\text{low}}$ represents the separated low-frequency information, while $F_{\text{high}}$ represents the separated high-frequency information.

\begin{figure}
    \centering
    \includegraphics[width=1\linewidth]{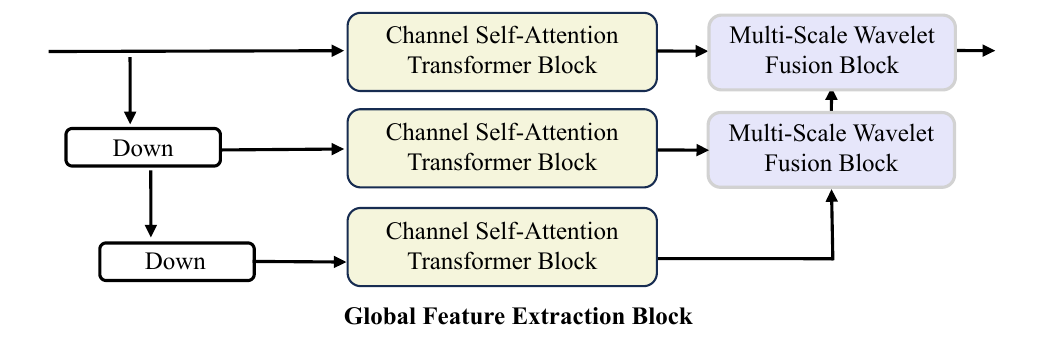}
    \caption{In the architecture of the proposed Global Feature Extraction Block, three successive downsampling operations are required. To prevent the loss of structural information during downsampling, the module employs cross-scale feature fusion based on wavelet transform.}
    \label{fig:GF}
    \vspace{-1.5em}
\end{figure}

As high-frequency information represents the details of the image, we need to adopt smaller receptive fields to better focus on local image information for finer detail restoration. To address high-frequency information, we propose the Local Feature Extraction Block for feature extraction, comprising convolutions with multiple small convolution kernels and dense connections. Small convolution kernels allow for better focus on detail areas, while residual connections excel at exploring high-frequency information\cite{dong2023multi,kim2016accurate}. Thus, this combination is highly effective for extracting high-frequency information.

For the low-frequency information of images, it contributes to the restoration of image backgrounds and outlines. Given that background and outline information occupies a considerable proportion of the image, having long-distance dependencies is beneficial for their restoration. Therefore, as shown in Fig.~\ref{fig:GF}, we introduce the Global Feature Extraction Block. In this module, we employ multi-scale feature fusion to consider long-distance interactions, and utilize Transformer during feature learning to establish global contextual relationships. Taking inspiration from \cite{zamir2022restormer}, the Transformer architecture implemented here eschews spatial self-attention in favor of channel-wise self-attention. This is because spatial self-attention would incur an unacceptable computational burden. 

In addition, although multi-scale feature extraction can achieve long-range dependencies, structural information may be lost during the downsampling process. Inspired by the ability of wavelet transformation to model scale information in images\cite{huang2022winnet}, we propose the Multi-Scale Wavelet Fusion Block for multi-scale information fusion. The Discrete Wavelet Transform separates large-scale feature information into \{HH, HL, LH, LL\}. We first fuse the original small-scale information with LL, and then use Inverse Discrete Wavelet Transform to fuse the merged information with \{HH, HL, LH\}. This approach helps avoid the loss of structural information when directly upsampling small-scale feature maps and merging them with large-scale feature maps. Overall, the Global Feature Extraction Block downsamples the input features three times, applies channel-wise self-attention on feature maps of different sizes, and finally utilizes the Multi-Scale Wavelet Fusion Block to merge features of different scales.

\subsection{Training Loss}
Training the network on tonemapped images is more effective than training directly in the HDR domain due to the common practice of displaying HDR images after tonemapping. Upon receiving an HDR image H in the HDR domain, we compress the image's range through the $\mu$-law transformation.
\begin{equation}
T(H) = \frac{\log(1 + \mu H)}{\log(1 + \mu)},
\end{equation}
where $\mu$ denotes a parameter that specifies the extent of compression, while $T(H)$ signifies the tonemapped image. Throughout our study, we constrain the values of $H$ to fall within the interval [0, 1], and we fix $\mu$ at 5000.
\begin{equation}
L_1 = \lVert T(H) - T(\hat{H}) \rVert_1,
\end{equation}
where $\hat{H}$ represents the predicted outcome derived from our HLNet model, while $H$ denotes the ground truth. In this method, we utilize the $L_1$ loss function to calculate the loss.
\section{Experiments}
\label{sec:Experiments}

\subsection{Experiments Settings}
\textbf{Datasets.}
\begin{table*}[t]
    \centering
    \small
    \setlength{\tabcolsep}{2pt} 
    \caption{The evaluation results on the Bracketing Image Restoration and Enhancement Challenge - Track 2 BracketIRE+ Task's dataset \cite{zhang2024bracketing}. We use NVIDIA RTX A100 GPU to calculate the inference time. The best and second best results are highlighted in \textbf{Bold} and \underline{Underline}, respectively.}
    \label{tab:quantitative results}
    \begin{tabular}{c|ccccccc}
       \toprule
        \textbf{Models} & \textbf{AHDRNet\cite{yan2019attention}} & \textbf{CA-ViT\cite{liu2022ghost}} & \textbf{XRestormer\cite{chen2023comparative}} & \textbf{TMRNet\cite{zhang2024bracketing}} & \textbf{Kim\cite{kim2023joint}} & \textbf{ESRT\cite{lu2022transformer}} & \textbf{Ours}  \\
        \midrule
        PSNR-$\mu${$\uparrow$} & 26.37 & 27.44 & 28.79 & 28.91 & 29.02 & \underline{29.11} & \textbf{29.66}   \\
        SSIM-$\mu${$\uparrow$} & 0.8479 & 0.8518 & 0.8566 & 0.8524 & 0.8571 & \underline{0.8579} & \textbf{0.8598}  \\
        Time(s) & 0.691 & 0.749 & 0.807 & \textbf{0.413} & 0.684 & 0.711 & \underline{0.642}  \\
        \#Params(M) & 17.65 & 17.90 & 17.88 & 13.58 & 18.07 & 18.12 & 17.60  \\
         
        \bottomrule
       
    \end{tabular}
\end{table*}
The dataset we utilized is the training set from the Bracketing Image Restoration and Enhancement Challenge - Track 2 BracketIRE+ Task. The data was obtained through a simulation process proposed by Zhang  \cite{zhang2024bracketing}. The dataset comprises a total of 1,335 data pairs from 35 scenes. Each data pair includes two inputs of different sizes, $\times$2 and $\times$4, with our task utilizing the $\times$4 size. Each input consists of five frames of raw images with different exposures, sized (4,135,240). In this dataset, 1,045 data pairs from 31 scenes were used for training, while the remaining 290 data pairs from four other scenes were reserved for testing.

\begin{figure*}[h]
  \centering

  \subfloat[]{%
    \includegraphics[width=1\linewidth, height=0.24\paperheight]{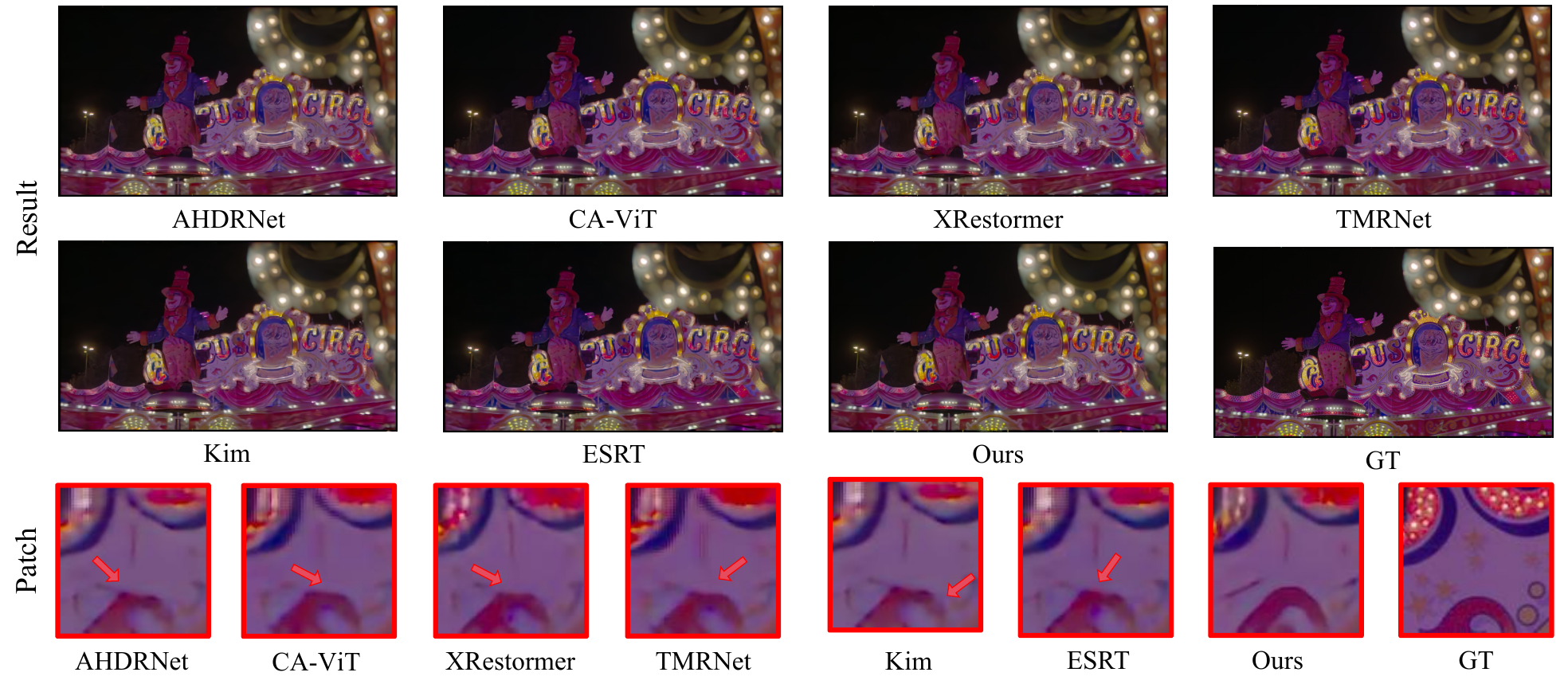}%
  }

  \hfill

  \subfloat[]{%
    \includegraphics[width=1\linewidth, height=0.24\paperheight]{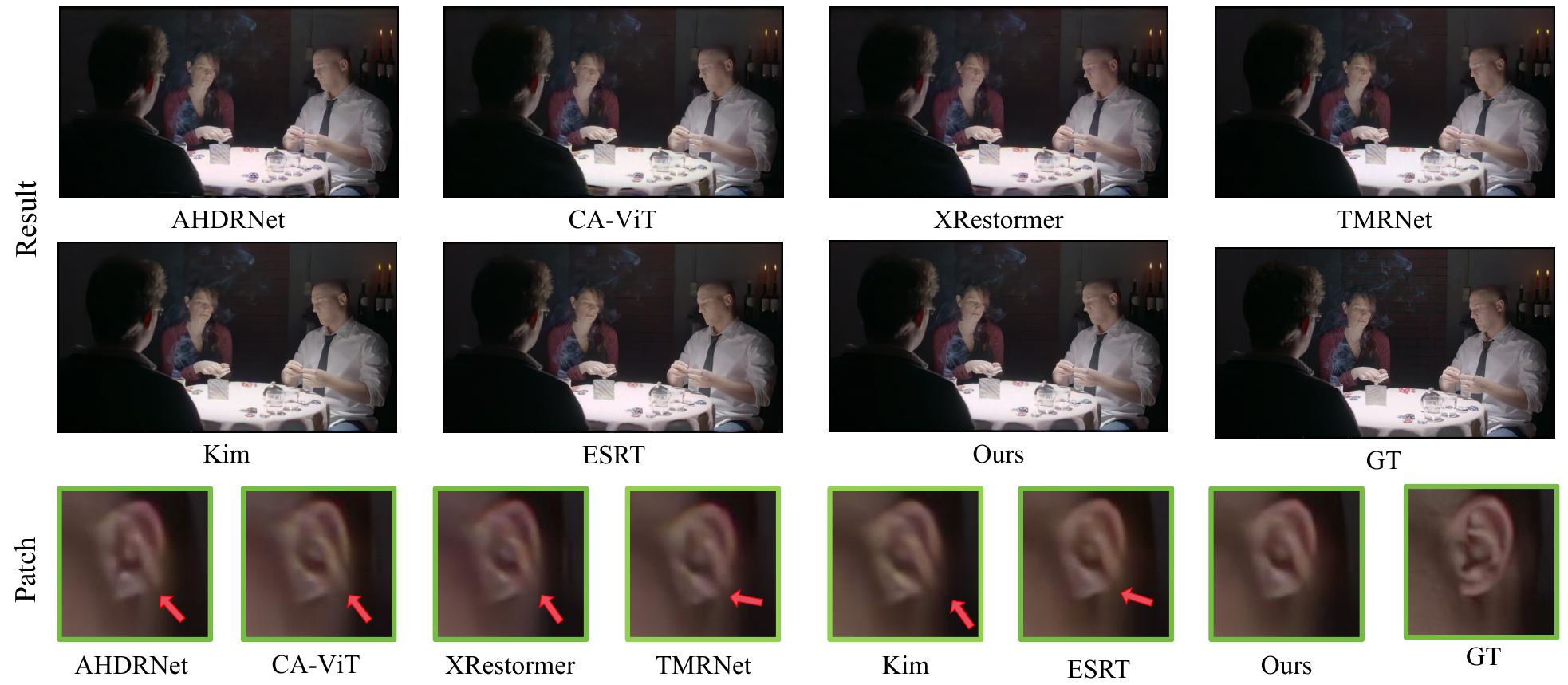}%
  }
  \hfill
  \subfloat[]{%
    \includegraphics[width=1\linewidth, height=0.24\paperheight]{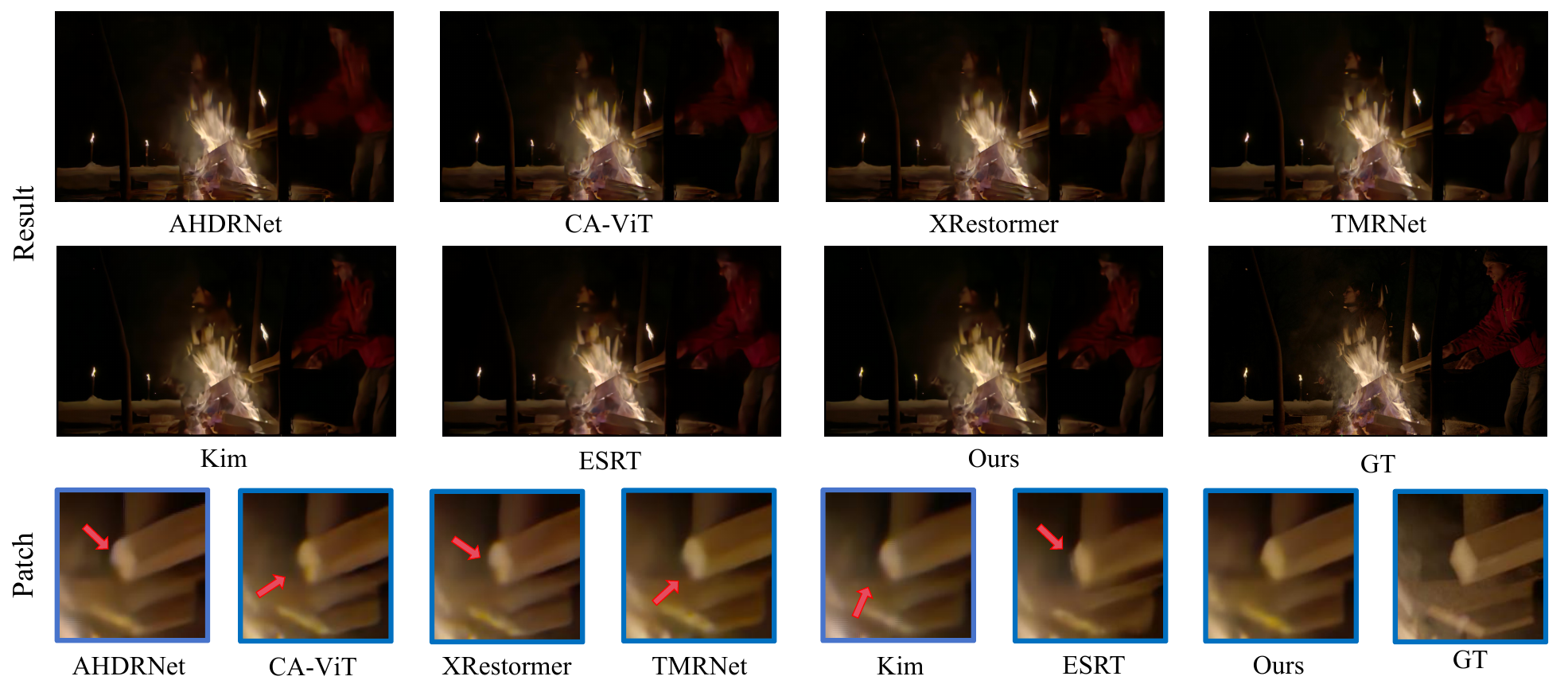}%
  }

  \caption{Examples of comparisions on the track 2 of the Bracketing Image Restoration and Enhancement Challenge dataset.}
  \label{fig:comparison}
\end{figure*}

\textbf{Evaluation Metrics.}
We use two objective measures for quantitative comparison: PSNR-$\mu$, SSIM-$\mu$ Here, $\mu$  indicate that the metrics are calculated in the tonemapped domain, respectively.

\textbf{Implementation Details.}
During the training process, both multiple downsampling operations and small input image sizes can affect training effectiveness. We cropped the images to a size of 64$\times$64 with a stride of 32, ensuring that the feature sizes after multiple downsampling steps were sufficient for effective feature extraction. We utilized the PyTorch framework and employed the AdamW optimizer with $\beta_1$ = 0.9 and $\beta_2$ = 0.999. Training was conducted on a single A100 GPU using the synthetic dataset provided by the BracketIRE+ Task for a total of 200 epochs, requiring 6 days of training.

\subsection{Comparison with the State-of-the-art Methods}

To validate the superiority of our model, we compared it with three HDR reconstruction models, two super-resolution models, and one unifying image restoration and enhancement model. These models' tasks are closely related to ours. The HDR reconstruction models, AHDRNet\cite{yan2019attention}, CA-ViT\cite{liu2022ghost}, and Kim\cite{kim2023joint}, reconstruct HDR images using multiple LDR frames, similar to our task. The super-resolution models, XRestormer\cite{chen2023comparative} and ESRT\cite{lu2022transformer}, are currently among the best-performing super-resolution models. The unifying image restoration and enhancement method's model, TMRNet\cite{zhang2024bracketing}, is consistent with our task. To eliminate the influence of parameter differences, we set the model parameters of these six methods to be roughly the same.

For the three HDR reconstruction methods, we only changed the number of input images, and the model used its original alignment and feature extraction methods, and finally added upsampling to increase the image resolution. For the two super-resolution methods, we only used its feature fusion method in the feature fusion stage, and the alignment method and upsampling were consistent with our model.

As shown in Table \ref{tab:quantitative results}, our method leads the second-best by 0.55 dB in terms of PSNR-$\mu$. Overall, as depicted in \cref{fig:comparison}(a), the images restored by our method exhibit the best visual effect to the human eye. In terms of detail recovery, our model outperforms others, effectively restoring the original shapes while other models exhibit significant blurriness. In \cref{fig:comparison}(b), our method still performs the best in detail recovery. We can observe rich details in the ears of our images, while ears restored by other methods appear blurry. Furthermore, ESRT's results are commendable, attributed to its adoption of high-low frequency decomposition methods, although slightly inferior to our model's performance. In Figure \cref{fig:comparison}(c), only our method does not exhibit blurriness in the edge details of the stick, while all other methods do.
\subsection{Ablation Studies}

To validate the effectiveness of each component in our model, we conducted ablation experiments on the Track 2 BracketIRE+ Task dataset. We designed four different ablation experiments to assess the importance of different components, including: (1) Removing the Spatial-Channel Enhancement Block(SCEB) and replacing it with a simple residual block; (2) Removing the High-Low Frequency Decomposition Block(HLFDB) and replacing it with a simple residual block; (3) Modifying the processing method of high-low frequency information in the High-Low Frequency Decomposition Block; (4) Modify the method of decomposing high-low frequency information. Consequently, a total of 6 different models were generated. The specific details of the models will be elaborated below, and the results of the ablation experiments are shown in Table \ref{tab:Ablation for HLNet}.

\begin{itemize}
\item \textbf{Model1:}  In the shared-weight module, we replaced the SCEB module with a simple residual block, and this model was named HLNet-NoSCEB.
\item \textbf{Model2:} In the non-shared-weight module, we replaced the HLFDB module with a simple residual block, and this model was named HLNet-NoHLFDB.
\item \textbf{Model3:} In the HLFDB module, when processing high-low frequency information, we used the Local Feature Extraction Block for both, and this model was named HLNet-LL.
\item \textbf{Model4:} In the HLFDB module, when processing high-low frequency information, we used the Global Feature Extraction Block for both, and this model was named HLNet-GG.
\item \textbf{Model5:} In the HLFDB module, when processing high-low frequency information, we use the method of processing high-low frequency information in ESRT to process high-frequency features and low-frequency features, and this model was named HLNet-ESRT.
\item \textbf{Model6:} In the HLFDB module, we replaced the previous method of decomposing high and low-frequency information using Average Pooling with the wavelet transform method, and this model was named HLNet-Wavelet.
\end{itemize}

\textbf{Without Spatial-Channel Enhancement Block.}
To validate the effectiveness of the Spatial-Channel Enhancement Block, we replaced it with a regular residual block. By referring to Table \ref{tab:Ablation for HLNet}, we observed a decrease of 0.25dB in the PSNR-$\mu$ metric compared to our model. Based on this result, we conclude that SCConv, by considering the relationship between spatial and channel information, can capture more comprehensive contextual information and dependencies between channels. The ability to capture contextual information and channel dependencies is highly beneficial for extracting common features across different degradations.
\begin{table}[t]
\caption{The Ablation study of HLNet on the Track 2 BracketIRE+ Task dataset.}
\centering
\begin{tabular}{l|cc}
\hline
\textbf{Models} & \textbf{PSNR-$\mu$} & \textbf{SSIM-$\mu$} \\

\hline
HLNet-NoSCEB & 29.41 & 0.8581 \\
\hline
HLNet-NoHLFDB & 28.97 & 0.8560 \\
\hline
HLNet-LL & 28.69 & 0.8521 \\
\hline
HLNet-GG & 28.75 & 0.8547 \\
\hline
HLNet-ESRT & 29.27 & 0.8566 \\
\hline
HLNet-Wavelet & 29.32 & 0.8577 \\
\hline
Ours & \textbf{29.66} & \textbf{0.8598} \\
\hline
\end{tabular}
\label{tab:Ablation for HLNet}

\end{table}

\begin{figure}[t]
    \centering
    \includegraphics[width=1\linewidth]{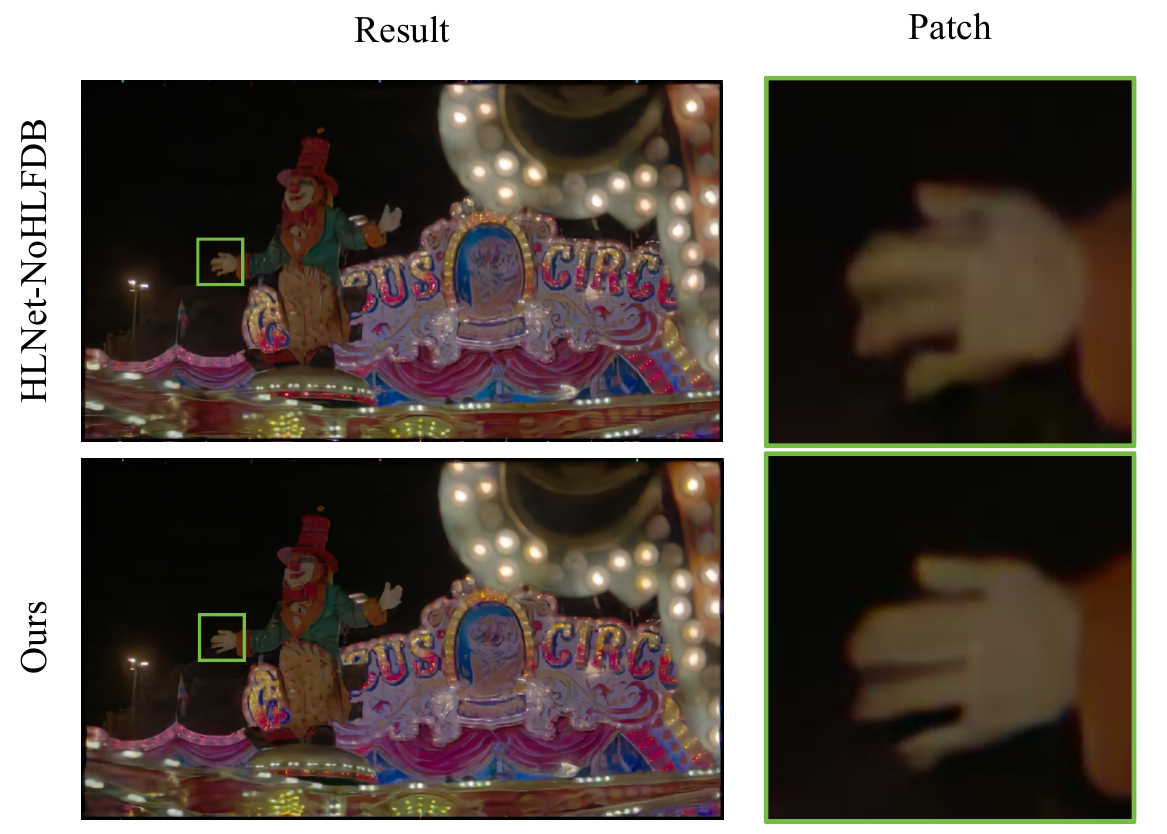}
    \caption{From the figure, it is evident the importance of the High-Low Frequency Decomposition Block. Models without using HLFDB show poor detail recovery. HLNet-NoHLFDB represents the absence of HLFDB in our model.}
    \label{fig:FD}
    \vspace{-1.0em}
\end{figure}

\textbf{Without High-Low Frequency Decomposition Block.}
To validate the effectiveness of the High-Low Frequency Decomposition Block, we replaced it with a regular residual block. By referring to Table \ref{tab:Ablation for HLNet}, we observed that this module is crucial, as removing it resulted in a decrease of 0.69dB in the PSNR-$\mu$ metric, which is unacceptable. This further confirms the importance of this module. Additionally, we can also perceive the advantages of this module in detail restoration visually. As shown in Fig.~\ref{fig:FD}, removing the High-Low Frequency Decomposition Block significantly reduces the details in the resulting image. There is severe blurring in the hand area. However, with the High-Low Frequency Decomposition Block, the hand area is well restored.

\textbf{Altering the approach to processing high-low frequency information in the HLFDB module.}
To validate the rationality and effectiveness of processing high-frequency information with convolution and low-frequency information with Transformer, we designed three sets of experiments: in the first set, both high-frequency and low-frequency information were processed with convolution; in the second set, both high-frequency and low-frequency information were processed with Transformer; in the third set, we used the method of processing high-low frequency information in ESRT to process features. From Table \ref{tab:Ablation for HLNet}, we observed that all three methods yielded poor results. Therefore, we conclude that it is reasonable and effective to process them with different methods according to the different characteristics of high-low frequency information.

\textbf{Using wavelet transform to decompose high-low frequency information.}
To validate the effectiveness of using average pooling for decomposing high and low-frequency information, we conducted an experiment using wavelet transform to decompose high and low-frequency information. From Table \ref{tab:Ablation for HLNet}, we can see that although the method of using wavelet transform to decompose high and low-frequency information does not perform as well as using average pooling for decomposition. This is because wavelet transform, when inappropriate wavelet basis functions are chosen, may lead to the loss of important details in the image, affecting the final image quality.

\subsection{Results of NTIRE 2024 Challenge on Bracketing Image Restoration and Enhancement - Track 2 BracketIRE+ Task}
\begin{table}[t!]
    \centering
    
    \footnotesize
    \caption{Results on track 2 of the Bracketing Image Restoration and Enhancement Challenge\cite{zhang2024ntirebrack}.}
    \vspace{-1mm}
    \label{tab:track2results}
    \begin{tabular}{cccc}
    \toprule
    Rank & Team & PSNR $\uparrow$ / SSIM $\uparrow$ / LPIPS $\downarrow$ & \#Params(M) \\  
    \midrule
    1 & SRC-B & 34.26 / 0.8913 / 0.206 &95.00  \\
    2 & NWPU & 30.59 / 0.8728 / 0.268 &13.37  \\
    3 & FZU\_DXW  & 29.82 / 0.8537 / 0.282 &14.34  \\ 
    4 & CYD & 29.66 / 0.8598 / 0.284 &17.60  \\
    5 & CVG  & 29.25 / 0.8521 / 0.278 &71.82  \\
    \midrule
    - & TMRNet~\cite{zhang2024bracketing} & 28.91 / 0.8572 / 0.273 & 13.58  \\
    \bottomrule
    \end{tabular}
    \vspace{-1.0em}
\end{table}
We participated in the Bracketing Image Restoration and Enhancement Challenge - Track 2 BracketIRE+ Task at NITRE2024 and achieved the fourth place. The results are shown in Table \ref{tab:track2results}. Our PSNR score is 0.75dB higher than the baseline model TMRNet.

\section{Conclusion}
\label{sec:Conclusion}
In this paper, we propose an approach called HLNet, which is based on high-low frequency decomposition for Bracketing Image Restoration and Enhancement. Typically, image restoration requires low-frequency information, while image enhancement requires high-frequency information. Therefore, to uniformly address different degradation issues, we propose a method based on high-low frequency decomposition, which can simultaneously provide the high-frequency and low-frequency information required for degraded images.

{
    \small
    \bibliographystyle{ieeenat_fullname}
    \bibliography{main}
}
\end{document}